\documentclass[conference]{IEEEtran}
\IEEEoverridecommandlockouts
\usepackage{graphicx}
\usepackage{subfigure}
\usepackage[ruled]{algorithm2e}
\usepackage{verbatim}
\usepackage{amsmath}
\newtheorem{mydef}{Definition}

\begin{document}
\title{PS-Sim: A Framework for Scalable Simulation of Participatory Sensing Data}
\author{\IEEEauthorblockN{Rajesh P Barnwal\IEEEauthorrefmark{1}\IEEEauthorrefmark{3}, Nirnay Ghosh\IEEEauthorrefmark{2}, Soumya K Ghosh\IEEEauthorrefmark{3}, and Sajal K Das\IEEEauthorrefmark{4}}\\
\IEEEauthorblockA{\IEEEauthorrefmark{1}IT Group, CSIR-Central Mechanical Engineering Research Institute, Durgapur, India (r\_barnwal@cmeri.res.in)}
\IEEEauthorblockA{\IEEEauthorrefmark{2}Department of CSE, Indian Institute of Information Technology, Kalyani, India (nirnay@iiitkalyani.ac.in)}
\IEEEauthorblockA{\IEEEauthorrefmark{3}Department of CSE, Indian Institute of Technology Kharagpur, India (skg@iitkgp.ac.in)}
\IEEEauthorblockA{\IEEEauthorrefmark{4}Department of CS, Missouri University of Science and Technology, Rolla, MO, USA (sdas@mst.edu)}}

\maketitle

\begin{abstract}
Emergence of smartphone and the participatory sensing (PS) paradigm have paved the way for a new variant of pervasive computing. In PS, human user performs sensing tasks and generates notifications, typically in lieu of incentives. These notifications are real-time, large-volume, and multi-modal, which are eventually fused by the PS platform to generate a summary. One major limitation with PS is the sparsity of notifications owing to lack of active participation, thus inhibiting large scale real-life experiments for the research community. On the flip side, research community always needs ground truth to validate the efficacy of the proposed models and algorithms. Most of the PS applications involve human mobility and report generation following sensing of any event of interest in the adjacent environment. This work is an attempt to study and empirically model human participation behavior and event occurrence distributions through development of a location-sensitive data simulation framework, called \emph{PS-Sim}. From extensive experiments it has been observed that the synthetic data generated by \emph{PS-Sim} replicates real participation and event occurrence behaviors in PS applications, which may be considered for validation purpose in absence of the ground-truth. As a proof-of-concept, we have used real-life dataset from a vehicular traffic management application to train the models in \emph{PS-Sim} and cross-validated the simulated data with other parts of the same dataset.
\end{abstract}
\begin{IEEEkeywords}
Participatory sensing, Human participation, Event reporting, Simulation framework
\end{IEEEkeywords}
\IEEEpeerreviewmaketitle
\section{Introduction}\label{intro}
During recent times, overwhelming development in the smartphone technology and its ubiquitous nature have lead to rolling out a plethora of participatory sensing (PS)~\cite{burke06} applications to help a user in making informed decisions. In PS, a human user, equipped with smart hand-held device (viz., smartphone, tablet, etc.) acts as a sensing agent typically in exchange of monetary or entertainment/educational incentives. If the user perceives anything worth sharing in his local environment, he submits the sensed information to a central server, which combines such information to build a body of knowledge. 

The PS data can have different modalities viz., text, video, audio-clip, image, and so on. Also, these data are generated in real-time based on sensing of events of interest, and in most cases explicitly reveal mobility traces. Furthermore, mining of such data gives interesting insights into human mobility, urban planning, necessity of civic services, and so on. Additionally, PS reduces infrastructure cost required to deploy and maintain sensors for data collection. A number of avenues such as traffic and road condition monitoring, air and noise pollution, gas pricing, waste and litter management, and so on have been found to adopt PS paradigm to collect user feedbacks either to improve their services as a part of the feedback loop, or publish information which are eventually beneficial to the society at large. It is to be noted that the mobility of humans plays an important role in sensing the environment. Thus, for our work we assume those PS applications which requires mobility for collection of environmental data. Given the benefits PS has over society and economy, few attempts to develop a more reliable and cost-effective PS platforms have recently been made~\cite{wang13}~\cite{wang15}~\cite{barnwal16}.

One major obstacle in PS research is the sparsity of the sensed data. To counter this, large scale deployment of test beds, development of middlewares, design of user interface, recruitment of volunteers for sensing tasks, and provisioning of incentives are required~\cite{gao15}~\cite{restuccia16}. However, this process of data generation is costly in terms of both time and money~\cite{cardone14}. Often the volunteers get demotivated as the sensing task consumes time and resources (e.g., battery energy, cellular data, etc.), resulting in the submission of irregular and unreliable data. Moreover, the perennial problem of inaccessibility of data generated by proprietary/commercial applications still persists. They either do not share data or allow limited exposure of their datasets in the public domain. On the flip side, the research community needs sizeable ground-truth data for validating their proposed models and algorithm.

Henceforth, short-term real PS datasets available in the public domain are the only source of information that can act as the representative data to study the efficacy of new proposals. Alternatively, there is a need to develop and use simulation framework which can emulate the participation behavior of real participants and can generate spatio-temporal distributions of the events and notifications. It is to be noted that such simulator is expected to be sensitive towards spatio-temporal biases and one needs to tune its parameters to generate synthetic data for different locations and time windows. Moreover, the simulator should scale with the size of participants and event reports, and also address various implementation related issues like incentive planning, recruitment of optimal number of participants etc. 

Simulation of PS data which mostly results out of human mobility requires understanding of the participation and event behaviors in real world. In~\cite{gedeon15}, a spatio-temporal coverage optimization problem in participatory sensing system has been designed and the solution reflects the effect of various mobility characteristics on such coverage. Zhang \emph{et al.}~\cite{zhang12} presents a spatio-temporal manifold learning algorithm to study the correlation of different urban sensing physical process. The work also discussed the traffic density variations during 24 hours time span. In~\cite{foell13}, a real dataset has been analyzed to predict the future transport usage through formulation of the probability density function of week-based number of travel days per user and probability mass function of travel periodicity across all users. Reddy \emph{et al.}~\cite{reddy08} proposed a set of metrics to determine the fitness of individual participant in the sensing task. In~\cite{tomasini17}, the authors studied the effect of human mobility on mobile sensor networks.

It is clear from the existing work that few efforts have been made in the past by the researchers to study the distribution of urban sensing process by humans, modeling of human mobility, prediction of transport usage and effect of human mobility on the mobile sensor networks. To the best of our knowledge, study of the participation behavior of users in terms of their contributions to the PS platform, and then extending this study to develop a data simulation framework has not been attempted. This limitation in the state-of-the-art PS research forms the motivation for our work.

In this work, we model participation behavior of the users and event occurrence pattern and develop a location-sensitive simulator framework, \emph{PS-Sim}, to cater the requirements of ground-truth data for research experiments. Additionally, in order to deal with real-time PS data, we parallelize \emph{PS-Sim} framework through implementation of the \emph{MapReduce} algorithm running on a Hadoop cluster. Two main benefits which the framework offers are: (i) support for modeling and simulation of human participation and event generation behaviors to emulate PS mechanism, and (ii) easy scaling up or down to generate datasets of any duration and population size in polynomial time.

We consider a popular but proprietary vehicular traffic management smartphone application, called \emph{Waze}\footnote{www.waze.com}, as a proof-of-concept to validate the data generated by \emph{PS-Sim}. It is capable of modeling participation behavior and temporal distribution of reports in vehicular PS applications. The design principle of \emph{PS-Sim} is such that it depends on the spatial and temporal parameters for generation of mobility-based PS data. 

Thus, the major contributions of this work are: (i) study of real-world PS data (Waze) for knowledge discovery on users' participation patterns and temporal distributions of report generations; (ii) extract distribution function from the gathered knowledge to model participation behavior and temporal factor related to reported events; (iii) development of a simulator framework for scalable simulation and generation of the simulated data; and (iv) cross-validation of the synthetic data using parts of the Waze. The results show that with respect to temporal variations, the simulated dataset closely follows the distribution of the real data. Further, the simulation time analysis experiments show that \emph{PS-Sim} is scalable in terms of number of participants and duration of participations.  

The rest of the paper is organized as follows. Section \ref{sysmod} describes the system model. Section \ref{train} studies vehicular PS application dataset from different perspectives. Section \ref{PS-Sim} presents the architecture, models, algorithm, and implementation details relevant to the \emph{PS-Sim} framework. Section \ref{result} evaluates the framework through validation and scalability experiments. Section \ref{conc} draws conclusions and gives future research directions.

\section{System Model} \label{sysmod}
\begin{figure}[!htbp]
\begin{center}
\includegraphics[scale=0.5]{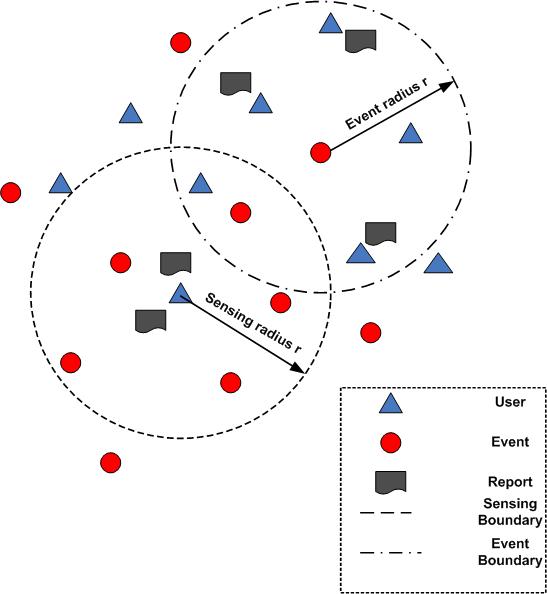}
\caption{System Model for PS \label{figsys}}
\end{center}
\end{figure}

As depicted in Fig.~\ref{figsys}, our system model captures a sensing region with $U$ users and $E$ events. All users are assumed to be equipped with smart devices and subscribed to a vehicular PS application which provides the interface to submit the sensed data. All the smart device are expected to have Wi-Fi or cellular data connectivity.

Three important components of this model are:\\
\underline{\em Report}: A report is an alert or notification generated by a user in response to his perception of an incident (viz., accident, jam, road closure, and so on).\\
\underline{\em Event}: An event is a notification which is published in some form (e.g., live map notice) after the PS platform receives a pre-defined number of ``similar'' reports from the users. Formally a PS event is defined as follows~\cite{barnwal16}:
\begin{mydef} \label{event}
(PS event). A PS event is defined as a four-tuple $Event$ = $\langle Date, DayTime, Loc, IncidentType\rangle$, where $Date$ denotes the date of occurrence of the event, $DayTime$ is the time-slot of the day when the event took place, $Loc$ is the location where the event has happened, and $IncidentType$ is the type of incident occurred.
\end{mydef}
\underline{\em User}: A user $u_i \in U$ is a participant in the PS application. $u_i$ senses his environment and has the propensity to generate reports, if he perceives anything of interest.

Modeling of participation and event occurrence behaviors need consideration of spatio-temporal biases. In our model, temporal biasness of event distribution is realized in terms of \emph{time of the day} and \emph{day of the week}. The underlying assumption is that, under normal condition, events occurring at the same time interval on weekdays receives similar number of reports, while those generated during weekends are different.

We assume each event to have a fixed perimeter known as the \emph{event boundary}. Within a particular event boundary, multiple users may be present. The users included in an event boundary are liable to generate reports. Similarly, each user also has a fixed \emph{sensing boundary}. In a given sensing boundary, multiple events of interest may take place simultaneously. The user has the obligation to report all or part of the events occurring in his vicinity. Through such considerations, we can divide the dataset into multiple regions, with the assumption that locations belonging to the same region has similar event occurrence pattern.

\section{Observations from the Data} \label{train}
We have used real-world dataset available from \emph{Waze}\footnote{https://data.cityofboston.gov/} as the ground-truth to train our participation and event generation models. The data comprises of traffic incident alerts generated between \emph{23-February-2015} and \emph{1-March-2015} by the users of the application. It has approximately 22,910 users, 71,505 reports for various traffic events occurring in 991 streets (locations) across 11 boroughs of Massachusetts (MA).

\begin{figure*}[!htbp]
\begin{center}
\subfigure[Report Generation Frequency\label{repfreq}]{
\includegraphics[scale=0.45]{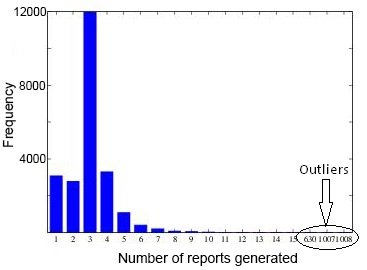}}
\subfigure[Report Distribution: Average Per Day\label{dayfreq}]{
\includegraphics[scale=0.2]{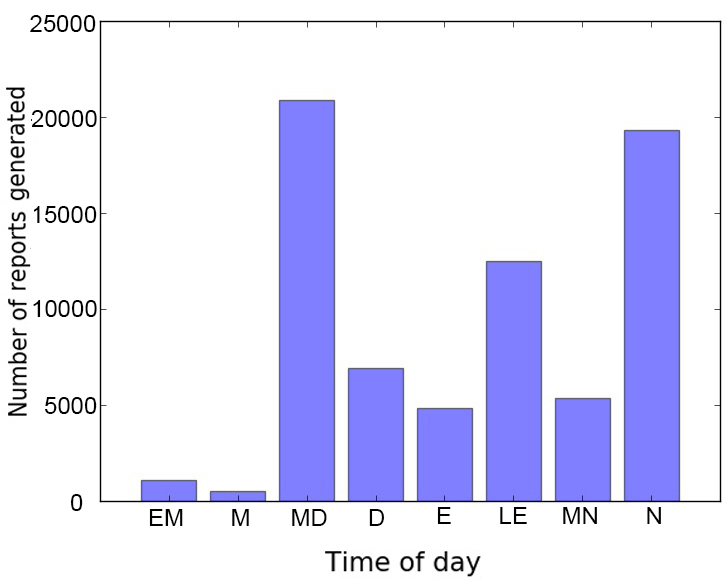}}
\subfigure[Report Distribution: Per Week\label{wkfreq}]{
\includegraphics[scale=0.2]{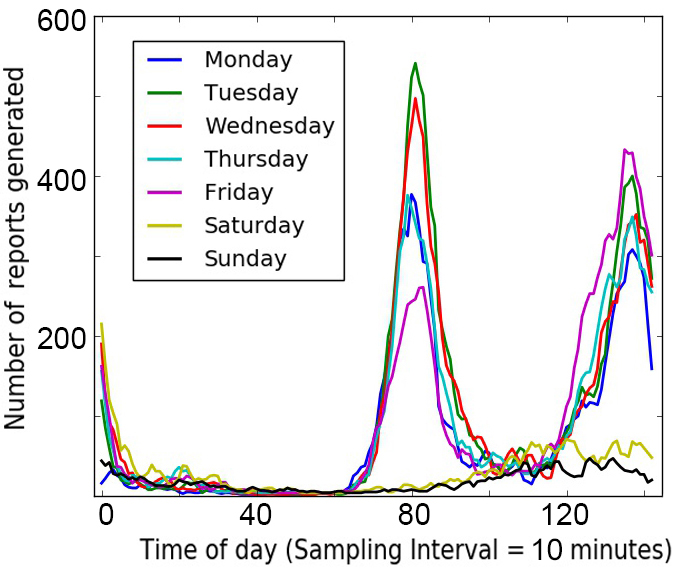}}
\caption{Findings from Waze Data \protect[Data Source: https://data.cityofboston.gov]}
\end{center}
\end{figure*}
The Waze dataset has been analyzed from various perspectives which has yielded some interesting findings. Considering the number of reports generated to be a random variable ($N = 1, 2, \ldots$), Fig.~\ref{repfreq} shows the number of users and their respective participation frequency in a week. It is evident that the majority of the users have generated around three reports over the span of a week except a few outliers.

One important feature we have selected from the dataset is the time (with respect to Coordinated Universal Time) a report got injected in the Waze database. To avoid redundant reporting, we have discretized the duration of a day into eight \emph{temporal bins}, each with a span of three hours starting at 3AM: \emph{early morning (EM), morning (M), day (D), mid-day (MD), evening (E), late evening (LE), mid-night (MN),} and \emph{night (N)}. Fig.~\ref{dayfreq} depicts the frequencies of report generation at different times over a day. It is evident that the distribution of reports is bimodal implying a higher percentage of human participation during two time windows over a day: \emph{mid-day} and \emph{night}. In general, less people commute during the weekend. This results into the lesser traffic events on the road and is responsible for generation of lower number of event reports. 

The generation of higher number of reports during \emph{mid-day}, \emph{late evening} and \emph{night} shows that those are the peak-hours of the day, which is obvious considering most businesses and entertainment activities take place during those time intervals. However, we also observe that few users have generated a large number of reports (between 600-1000). As can be seen from Fig.~\ref{repfreq}, these reports are mostly outliers, given the frequency at which they have been generated. This entails the requirements of pre-processing and cleaning of the dataset for filtering out the outliers and missing values.

Similar to \emph{time bins}, we also form \emph{day bins} representing the number of reports generated on each day of the week. Fig.~\ref{wkfreq} presents the frequency of generation of reports on different days of a week with a sampling rate of 10 minutes. From the plot, we can observe that during weekdays, i.e., Monday - Friday, the distribution of the event reports are following the different trends for different days. The peak participation size on Monday is closer to that of Thursday than of Friday. Interestingly, on Friday the peak is on higher side during evening than that during afternoon. And this creates uniqueness in Friday’s event report generation. This possible reason may be  due to forthcoming weekend, people are deferring their travels for the evening to take advantage of night entertainment and outing. Additionally, this has also been depicted from the day bins plot that the number of reports generated during Saturday and Sunday are relatively low.  

Based on these observations, we generate the \emph{probability mass functions (pmfs)} for the time bins and day bins. In the \textit{pmf}, the time bins and day bins are considered as random variables, and the fraction of report generated is the dependent variable. Considering $X$ to be the random variable, the \textit{pmf} for discretized temporal bins $f_{X}:A \rightarrow [0, 1]$, is defined as follows:
\begin{equation}\label{eqn:pmf}
P_{X}(x) = P(X = x) = Pr({s \in SS : X(s) = x})
\end{equation}
where $P_{X}(x)$ is the probability of happening $x$ number of events during temporal bin $s$ and $SS$ = \{Sunday, Monday, Tuesday, Wednesday, Thursday, Friday, Saturday\} or $SS$ = \{Early morning, Morning, Mid day, Day, Evening, Late evening, Mid night, Night\}.

We considered two different \emph{pmf}s for time bins and day bins. It is also ensured that Eqn.(\ref{eqn:03}) remains valid for applicability of $pmf$ theory.
\begin{equation}\label{eqn:03}
\sum_{x\epsilon A}f_{X}(x) = 1
\end{equation}
The consideration of \textit{pmf} for time or day bins provides an option for reconfigurability of the simulation as per the characteristics of the location under study.

\section{Proposed PS-Sim Framework} \label{PS-Sim}
In this section, we discuss the overall system architecture and the modeling aspects of human behavior and event occurrences, derived from the empirical study of the Waze dataset. Finally, we present the algorithm and the MapReduce implementation of the proposed simulator framework.

\subsection{Architecture} \label{PS-Sim:archi}
\begin{figure}[!htbp]
\begin{center}
\includegraphics[scale=0.5]{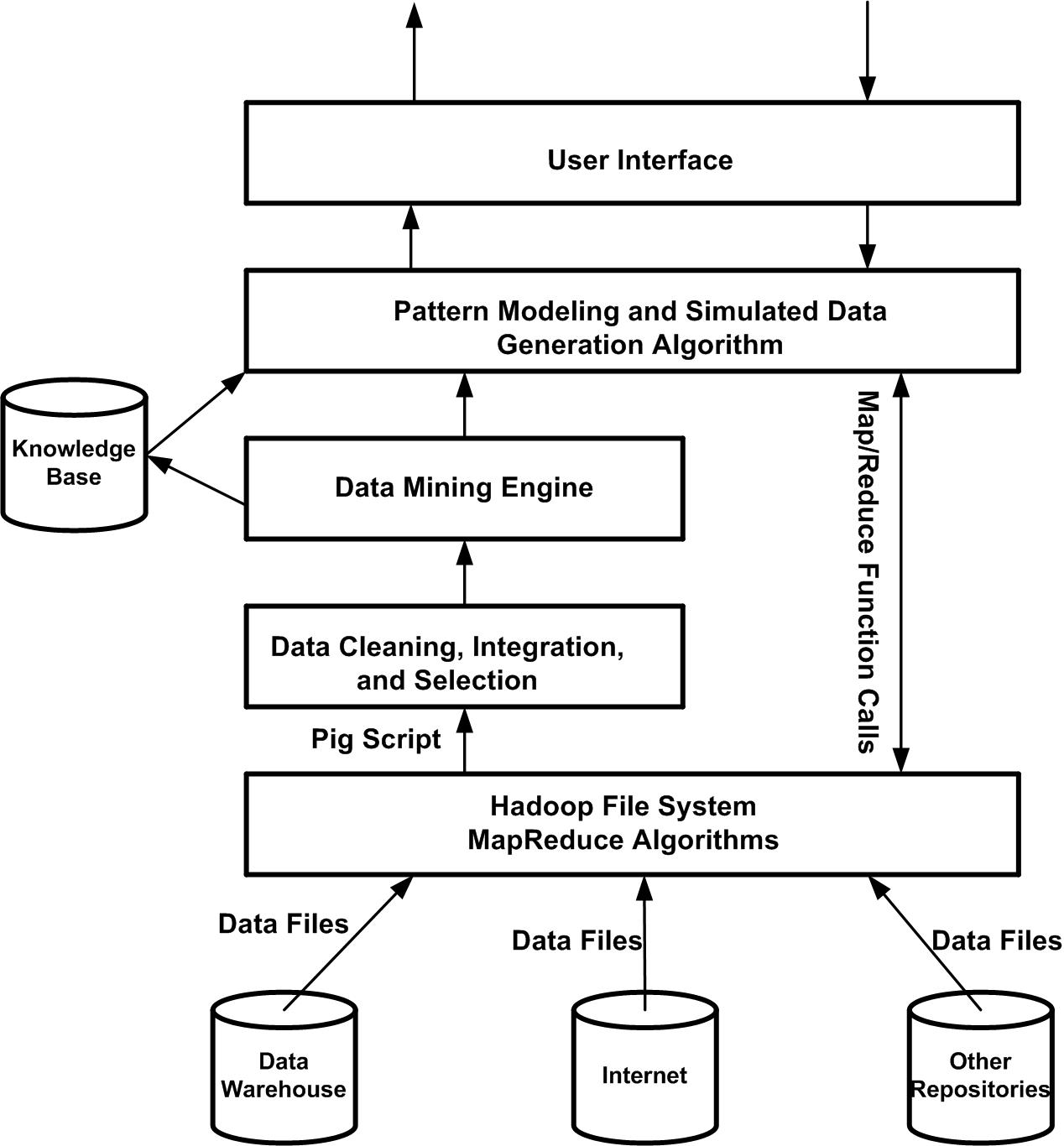}
\caption{PS-Sim Architecture}
\label{arch}
\end{center}
\vspace{-0.25in}
\end{figure}

The overall system architecture of the proposed \emph{PS-Sim} framework is depicted in Fig.~\ref{arch}. The implementation allows integration of datasets from different sources.

As evident from the architecture, PS data collected from various sources are given as input to the \emph{Hadoop File System (HDFS)} which has been deployed in a cluster running on our private cloud infrastructure. HDFS provides a framework for implementing the \emph{MapReduce} algorithms and facilitates distributed data processing across many nodes. We have leveraged its data processing and analysis tool, \emph{Apache Pig}, which utilizes the underlying MapReduce model to execute scripts. First, the stored data is filtered based on the requirements using \emph{Pig script}. Next, they are integrated and input to feature selection algorithms for appropriate feature selection. The selected feature set is then fed to the data mining engine (consisting of classifiers) to train our statistical models and store the labeled data into the knowledge base. The knowledge obtained from the training phase are given as input to the proposed algorithm to generate simulated (predicted) data.

In order to handle massive volumes of PS data, we extend our simulator framework in a parallel way, such that it can leverage \emph{Map} and \emph{Reduce} functionalities of the MapReduce algorithm to partition the dataset and perform parallel processing. The \emph{Map} function processes input values to generate a set of intermediate \emph{key/value pairs}, and the \emph{Reduce} function merges all intermediate values associated with the same intermediate key. Finally, the accuracy of the simulated dataset is tested using the 10-fold cross-validation technique.
\subsection{Modeling Participation Behavior} \label{PS-Sim:participate}
Modeling of human participation behavior is one of the key elements required for building the data simulation framework for PS applications. We argue that there exist a relationship between mobility and participation behavior of humans. Reports get generated only if participants are present at the vicinity of event, provided that no spoofing of locations or compromise of devices has taken place. Our objective is to model the distribution of the participants' contributions in terms of the number of reports generated per week. Thus, we partition the event report data according to spatial regions. We use 10-fold methods to split the region-based partitioned data for modeling and validation of the simulator output in terms of participation behavior and distribution of events. For analysis, the submitted reports have been grouped by participant.

\begin{figure}[!htbp]
\begin{center}
\includegraphics[scale=0.45]{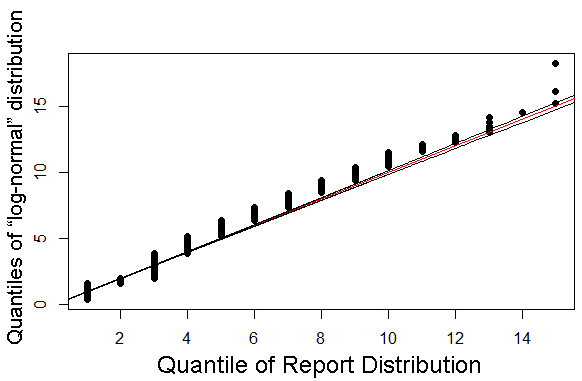}
\caption{Q-Q Plot: Participation}
\label{qq}
\end{center}
\end{figure}

Fig.~\ref{repfreq} shows the frequency of generation of reports. It can be found that the weekly contributions from majority of the participants are centered around a mean value and the plot is highly skewed towards left with a long tail towards right. To find the nearest distributions of the report generation frequency, we compared the data with different types of distributions using \textit{Q-Q Plot method}. A Q-Q (quantile-quantile) plot is a probability plot, which is a graphical method for comparing quantiles of one dataset against the quantiles of the second dataset. Through \emph{Q-Q Plot testing} it has been found that the data has strong correlation with \emph{log-normal distribution} (refer to Fig.~\ref{qq}). Moreover, this observation can be validated from the findings in~\cite{zhao14} which studied large datasets and concluded that human mobility follows log-normal distribution.

Thus, we modeled the participation behavior of users using log-normal distribution with $M$ as location parameter and $S$ as scale parameter. The probability of contributing $x_{p}$ number of reports by the participant $p$, corresponding to the events occurring within his sensing boundary is given by
\begin{equation}
\label{eqn:lognormal}
P(x_{p})=\frac{1}{Sx_{p}\sqrt{2\pi}}*e^{\frac{-(\ln{S}- M)^{2}}{2S^{2}}}
\end{equation}
The above participation model is scalable and can be scaled up or down by modifying the scale and location parameter according to the required time of the simulation. It is imperative that the model is adaptive to capture various contexts, viz. if there is some festival, or protest, or sports event, and so on. This can be achieved by progressive update of the scale and location parameters.

\subsection{Modeling Event Occurrence} \label{PS-Sim:event}
The likelihood of sensing an event and generating corresponding reports is a spatio-temporal process and depends upon two factors: (i) probability of occurrence of an event at different time slots, and (ii) distribution of events on different days of the week. Our initial hypothesis is that the distribution of the events and their reporting are time-dependent. For example, during peak hours of the day (morning, noon, and evening), there will be higher number of vehicles and humans on the streets (as evident from Fig.~\ref{dayfreq}). During those time, the odds of happening of an event at a particular location is always higher than that during non-peak hours. Similarly, barring some exceptions, the frequency of occurrences of events and their reporting will always be on the higher side during weekdays in comparison to that of during weekends (as seen in Fig.~\ref{wkfreq}). We assume that the generation of PS reports is proportional to the occurrence of the event.

\begin{figure}[!htbp]
\begin{center}
{
\includegraphics[scale=0.7]{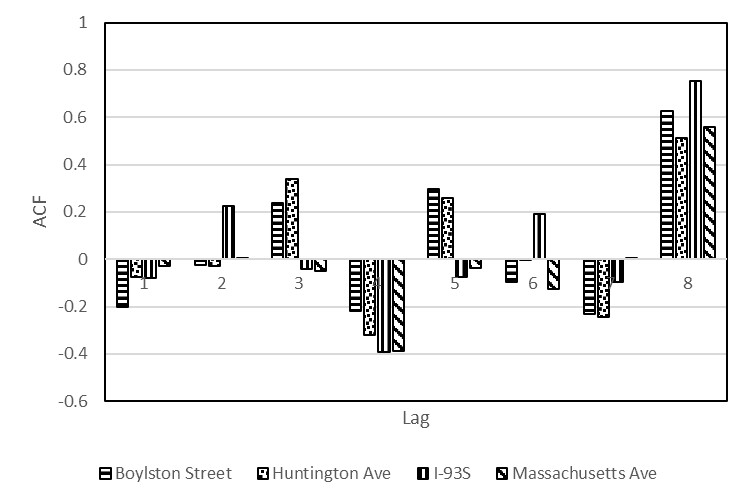}}
\caption{Autocorrelation: Street-wise Report Frequency}
\label{figauto}
\end{center}
\end{figure}

Furthermore, to discover the effects of the past event reports on the frequency of the forthcoming event reports, we experimented to find autocorrelation in the dataset. As mentioned earlier, we segregated the frequency of event reports in 56 temporal bins comprising of seven days with eight time bins in each day. Among 56 sets of data, we performed autocorrelation tests with lags starting from 1 to 8. From Fig.~\ref{figauto}, it is evident that the \textit{Auto-correlation Factor (ACF)} at different locations is very low for lags of 1 through 7. This reflects the fact that the frequency of event reports generation at present time is independent of that in past, except at the lag of 8, where the ACF is more than 0.6. It indicates that the event report frequency follows the similar trends and slight periodicity during same time bins of different days. Discarding exceptional cases, this is obvious as during every mid-night the event report frequency would be similar for each day and so is the true for the other time bins like, mid-day, evening etc. Furthermore, as shown in Table~\ref{tab:lambda}, the average rate of generating event reports varies for different streets which entails that each location has distinct characteristics and biases.
\begin{table}
\caption{Event Report Generation Frequency} \label{tab:lambda}
\vspace{-0.1in}
\begin{center}
\begin{tabular}{|l|c|}
\hline {\bf Street Name}&{\bf Rate of Event Reports}\\
\hline Boylston Street&$25.29$\\
\hline Huntington Ave&$10.55$\\
\hline I-93S&$38.39$\\
\hline Massachusetts Ave&$10.98$\\
\hline
\end{tabular}
\end{center}
\vspace{-0.25in}
\end{table}

The autocorrelation study provides an insight into the event report frequency and reveals that within a day, the occurrence of events and arrival of reports are independent between two time bins. The modeling of event distribution at a particular location is also an important consideration while developing the simulator framework. In the present work, we assume that the event happening at any time and in a particular location follows \emph{Poison point process} with rate $\lambda_{e}$ as shown in Eqn.(~\ref{eqn:pois}). This assumption gives us the opportunity to incorporate the fact that the occurrence of any event at any location is exponentially distributed and is independent of the past events~\cite{azadeh10}.
\begin{equation}
\label{eqn:pois}
P(k) = \frac{e^{-\lambda_{e}}\lambda_{e}^{k}}{k!}
\end{equation}
Where $P(k)$ is the probability of occurrence of $k$ events per temporal bin. The Poisson parameter $\lambda_{e}$ is the mean rate of generation of event reports and is characteristic of a particular location and varies with the type of location viz., Point of Interest (important road crossing, office place, shopping and entertainment area, etc.), city road, highway, or neighborhood.

We assume that each location has its own characteristics in terms of probability of happening of a particular type of event. For example, a particular street is more likely to experience traffic jam but the same street is less likely to be closed. Similarly, some street is more prone to accident whereas at some street like junction points, traffic jam is a more probable event. Such location-specific event type distribution can be done by varying the density parameter $\lambda_{e}$.
\subsection{Algorithm} \label{PS-Sim:algo}
\emph{PS-Sim} is a discrete time simulator that provides features to configure the various spatio-temporal biases that makes it more suitable for simulation of data generated in mobility-based PS applications. Table \ref{tab:parameter} describes the different configurable parameters. \emph{PS-Sim} simulates various events and their related PS reports in different discretized bins to produce realistic trends.
\begin{table}
\caption{Simulation Parameters} \label{tab:parameter}
\vspace{-0.1in}
\begin{tabular}{|l|l|}
\hline {\bf Parameter}&{\bf Description}\\
\hline $\tau$&Number of days\\
\hline $dt_{s}$&Starting date\\
\hline $evType[]$&List of event types\\
\hline $\Pr_{l}$&Probability that a participant lies during reporting\\
\hline $n$&Number of Participants\\
\hline $\lambda_{e}$&Rate of occurrence of events\\
\hline $mlog$&Average participation frequency per week\\
\hline $sdlog$&Standard deviation of participation frequency per week\\
\hline $\mu$&Rescaled average frequency of participation\\
\hline $\sigma$&Rescaled standard deviation in frequency of participation\\
\hline $tmBin[]$&Bin for times of a day\\
\hline $dyBin[]$&Bin for days of a week\\
\hline $pmf(tmBin)$&Probability distribution of events per temporal bin\\
\hline $pmf(dyBin)$&Probability distribution of events per day bin\\
\hline $pmf(evType)$&Probability distribution of event types\\
\hline
\end{tabular}
\vspace{-0.25in}
\end{table}
The working of \emph{PS-Sim} is described by Algorithm \ref{algo:01}. The simulator produces event reports simulating user participation patterns prevailing at a particular location. The algorithm first rescales the participation log-normal statistical parameters for the desired duration (that needs to be simulated). The value of statistical parameters are subjective and location-specific. This step helps to incorporate customized location-specific participation into the framework. Then, the algorithm uses the rescaled statistical parameters to generate participation frequency of individual participants. Further, the distribution of the events is modeled as Poisson point process with $dyBin$, $tmBin$ and $\lambda_{e}$ as the parameters. 

Next, the algorithm simulates event generations using probability mass functions for $dyBin[]$, $tmBin[]$ and $evType[]$. Since, PS systems are susceptible to false data contributions (may be due to perception difference or malicious intents of the participants), it is imperative that the simulator should also possess the feature for infusing false reporting scenario. To simulate the effect of the false report generation by the participants, a predefined probability factor, $Pr_l$, is used for generating random event reports in place of report for actually occurred events. The simulator finally generates a trace file comprising of participatory sensing event reports. Any event report $R$ is 8-tuple data structure comprising of $\langle EventNo$, $Date$, $Day$, $Time$, $ReportNo$, $SourceId$, $EventReported$, $EventOccured \rangle$. The sample reports generated by the simulator are shown in Table \ref{tab:data}.
\begin{table*}[ht]
\begin{center}
\caption{Sample Data as Generated using Simulator} \label{tab:data}
\begin{tabular}{|l|l|l|l|l|l|l|l|}
\hline {\bf EventNo}&{\bf Date}&{\bf Day}&{\bf Time}&{\bf ReportNo}&{\bf SourceId}&{\bf EventReported}&{\bf EventOccurred}\\
\hline 51&09/01/2016&Thursday&MidDay&112&UID000858&Accident&Jam\\
\hline 51&09/01/2016&Thursday&MidDay&119&UID000233&Jam&Jam\\
\hline 51&09/01/2016&Thursday&MidDay&134&UID000107&Jam&Jam\\
\hline 109&10/01/2016&Friday&Day&208&UID000632&Accident&Accident\\
\hline 109&10/01/2016&Friday&Day&232&UID000987&Jam&Accident\\
\hline 314&11/01/2016&Saturday&Morning&872&UID001323&Jam&Accident\\
\hline
\end{tabular}
\end{center}
\vspace{-0.2in}
\end{table*}
\begin{algorithm}
\caption{PS-Sim Simulator} \label{algo:01}
\SetAlgoLined \KwIn{$\tau$, $dt_{s}$, $evType[]$, $Pr_{l}$, $n$, $\lambda_{e}$, $mlog$, $sdlog$, $tmBin[]$, $dyBin[]$,$pmf(tmBin)$, $pmf(dyBin)$, $pmf(evType)$}
\KwOut{$R$}
\textbf{Initialize}: $EvDates$, $EvBin$, $tmBin$, $dyBin$, $\mu$, $\sigma$, \textit{N[]}, $c$,$R$\;
\nl $(\mu, \sigma) \leftarrow reScale(mlog, sdlog, \tau)$\;
\nl $Source[1..n] \leftarrow lognormal(n, \mu, \sigma)$\;
\nl $NoOfReports \leftarrow sum{N[1..n]}$\;
\nl $Events \leftarrow gen\_poisson\_ev(dyBin, tmBin, \lambda_{e})$ \;
\For{(\textit{i= 1 to Events})}
{
	\nl $EventNo[i] \leftarrow i$\;	
	\nl $EventDay[i] \leftarrow sample(dyBin[], 1, pmf(dyBin))$\;
	\nl $EventDate[i] \leftarrow sample(evDates[], 1)$\;
	\nl $EventTime[i] \leftarrow sample(tmBin[], 1, pmf(tmBin))$\;
	\nl $EventType[i] \leftarrow sample(evType[], 1, pmf(evType))$\;
}
\For {(\textit{repNo = 1 to NoOfReports})}
{
	\nl $evNo \leftarrow sample(EventNo, 1)$\;
	\nl $Id \leftarrow sample(1:n, 1)$ s.t. $Source[Id] > 0$\;
	\nl $EventOccured \leftarrow EventType[evNo]$\;
	\nl $EventReported \leftarrow EventOccured$\;
	\If{$Pr_l$}{
		\nl $EventReported \leftarrow sample(evType[]\backslash EventOccured, 1)$\;
}
$R \leftarrow R  \cup$\{\textit{EventNo, Date, Day, Time, ReportNo, SourceId, EventReported, EventOccured}\}
}
\textbf{Return}: $R$\;
\end{algorithm}

\subsection{MapReduce Implementation} \label{PS-Sim:mapreduce}
In our current implementation, we use the MapReduce functions to map the user generated reports to the corresponding events. From the \emph{Waze} dataset it is evident that multiple users generate one or more reports to notify a particular event. Each report contains the participant's ID, spatio-temporal information, and the type of incident perceived. Generation of simulated data needs each event to be in predefined representation (as given in Definition \ref{event}), and also a count of the number of reports supporting it.

In the mapping process, all the input reports are re-organized into key/value pair, where each key is the four-tuple $\langle Date, DayTime, Loc, IncidentType\rangle$ and the value is the rest of the information. Before these key/value pair is fed to Reducers, they will be sorted by Hadoop (according to temporal order) so that the pairs with the same key will go to the same Reducer. In the reducers, the number of occurrences of the respective key values are counted. Finally, for each event a key/value pair is generated, where the key is the four-tuple and the value is the number of supporting reports. Such parallel computation to find the frequency of event occurrences enables consideration of large input dataset for scalable generation of simulated results.

\section{Results and Discussion} \label{result}
\begin{figure*}[!htbp]
\begin{center}
\subfigure[Participation\label{fig:realSimPart}]{
\includegraphics[scale=0.2]{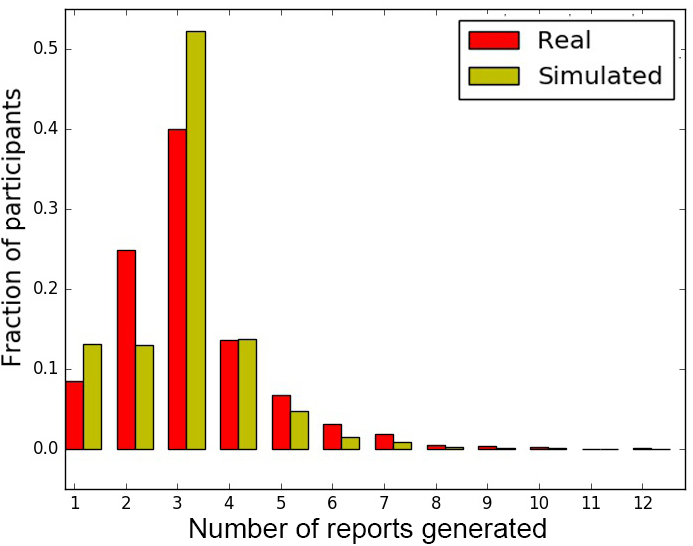}}
\subfigure[Weekly Report Generation\label{fig:realSimDay}]{
\includegraphics[scale=0.2]{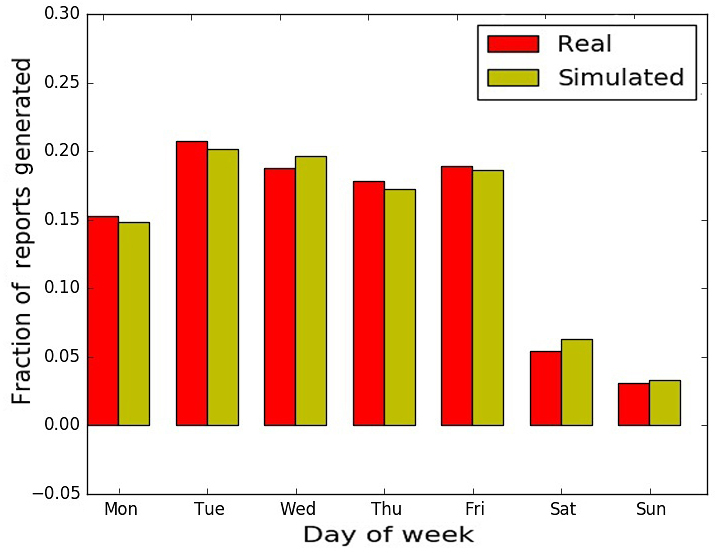}}
\subfigure[Daily Report Generation\label{fig:realSimTime}]{
\includegraphics[scale=0.2]{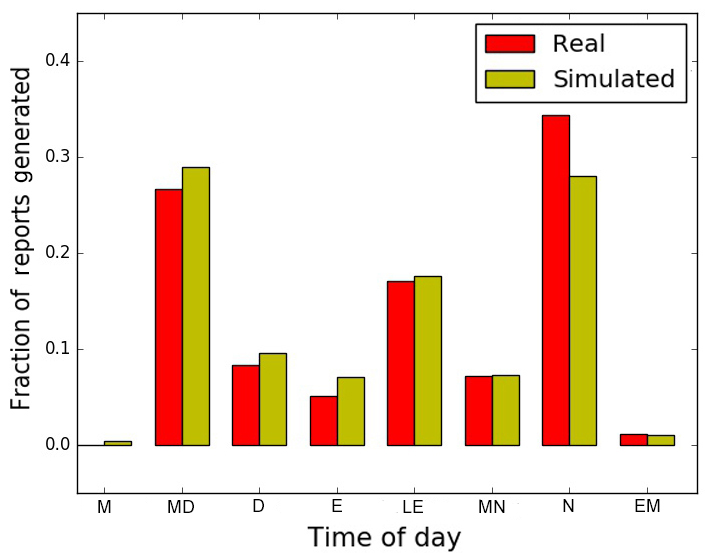}}
\caption{Comparison of Data: Real vs Simulated}
\end{center}
\end{figure*}

The working of simulator has been validated using 10-fold partitioning of the real dataset from \emph{Waze}. For each fold of test data, our main goal is to establish the similarity between the simulator output and the real data with respect to participation and event generation behaviors of the users. The distribution of real data and simulated data are plotted in Figs.~\ref{fig:realSimPart}, ~\ref{fig:realSimDay} and~\ref{fig:realSimTime} for comparison.
\begin{table}[ht!]
\caption{Correlation-RMSE Analysis} \label{tab:error}
\begin{center}
\vspace{-0.12in}
\begin{tabular}{|l|l|l|}
\hline {\bf Parameter}&{\bf Correlation}&{\bf RMSE}\\
\hline Reports per user&0.9389&0.0516\\
\hline Reports per day bin&0.9967&0.0058\\
\hline Reports per time bin&0.9791&0.0253\\
\hline
\end{tabular}
\end{center}
\vspace{-0.15in}
\end{table}
In Fig.~\ref{fig:realSimPart}, we consider the number of reports generated as a random variable and computed the fraction of users. It is evident that the participation behavior, in terms of event report contribution, is following similar distribution pattern for both real test data and the simulated data.  Also, Figs.~\ref{fig:realSimDay} and~\ref{fig:realSimTime} show that the temporal distribution of the event reports generated by the simulator are following same trends as the real test data. To check the quality of the simulated data, we made \emph{correlation analysis} and \emph{root-mean-square error (RMSE)} of the simulator generated data and the real test data. The results given in Table \ref{tab:error} show that the \textit{PS-Sim} framework generated data has high correlation and have low \textit{RMSE} when compared with the real test data for all three parameters: \emph{reports per user (participation frequency)}, \emph{reports per day bin}, and \emph{reports per time bin}.

\begin{figure}[!htbp]
\centering
\includegraphics[scale = 0.5]{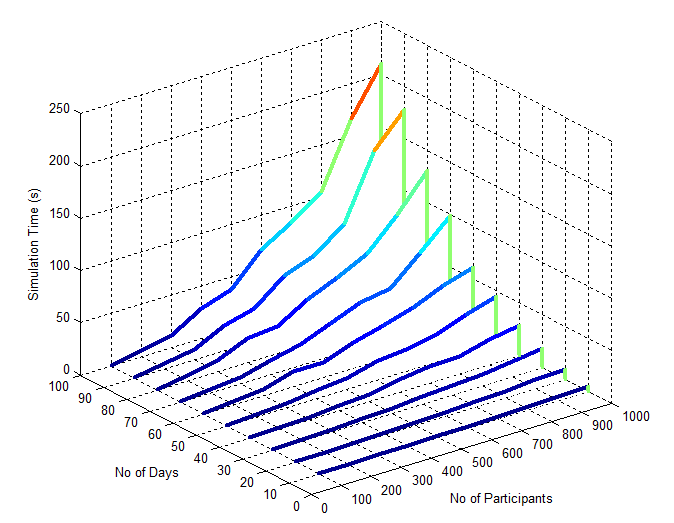}
\caption{Simulation Time Analysis}
\label{fig:simTimeAnalysis}
\vspace{-0.15in}
\end{figure}

To study the effect of MapReduce-based parallel computation on the scalability of \emph{PS-Sim}, we varied the number of participants $n$ and the number of days $m$ in the range $100\leq n \leq 1000$ and $10\leq m \leq 100$ respectively, where each user on an average generates three reports per week. The time taken to generate the simulated data has been plotted in Fig.~\ref{fig:simTimeAnalysis}.

It is evident that the simulation time is in the order of $200$ seconds for simulating the participation of $1000$ users and generating reports over a period of 100 days. Through curve-fitting analysis, we find the time complexity of our implementation is of the order of polynomial of power 2, i.e., $T(n,m)=O(n^{2}*m^{2})$. Moreover, due to the usage of multiple Mapper and Reducer functions, the simulator is capable of utilizing all the available CPU cores simultaneously. This helps to scale-up the simulator for larger scale data generation using higher core systems.
\vspace{-0.1in}
\section{Conclusion}\label{conc}
Mobile participatory sensing is an emerging field which produces huge volume of data and requires large scale test bed for research experiments. As human users are themselves sensing agents, persistent generation of truthful and up-to-date data is not always guaranteed. In this condition, the generation of near-optimal simulated data is one of the cost-effective and feasible alternatives. However, from the existing literature it is evident that there exists no such data simulator framework for PS applications, which can emulate the participation behavior and event occurrences in the realistic manner. In this paper, we propose a novel framework, \emph{PS-Sim}, which analyzes the PS data and models location-sensitive participation behavior of users in terms of their contributions and temporal activeness. It generates near-optimal simulated data which has been validated with real vehicular PS application dataset. The scalability of simulated data generation has been greatly improved by using MapReduce-based parallel computation and found to generate synthetic datasets of varying sizes in lower-order polynomial time. In future, we plan to extend our framework by incorporating more features viz. human mobility model and road traffic dynamics model, and reports of other modalities (viz. image, multimedia, etc.) to make it more comprehensive for use in PS research.


\begin{thebibliography}{13}
\scriptsize

\bibitem{burke06} J. A. Burke, D. Estrin, M. Hansen, A. Parker, N. Ramanathan, S. Reddy, and M. B. Srivastava. ``Participatory Sensing''. \emph{Center for Embedded Network Sensing}, 2006.
\bibitem{wang13} X. Wang, W. Cheng, P. Mohapatra, and T. Abdelzaher. ``ARTSense: Anonymous Reputation and  Trust in Participatory Sensing'', \emph{IEEE International Conference on Computer Communications (INFOCOM)}, pp. 2517-2525, 2013.
\bibitem{wang15} D. Wang, T. Abdelzaher, and L. Kaplan. ``Social Sensing: Building Reliable Systems on Unreliable Data''. \emph{Morgan Kaufmann}, 2015.
\bibitem{barnwal16} R. P. Barnwal, N. Ghosh, S. K. Ghosh, and S. K. Das. ``Enhancing Reliability of Vehicular Participatory Sensing Network: A Bayesian Approach''. \emph{IEEE International Conference on Smart Computing (SMARTCOMP)}, pp. 1-8, 2016.
\bibitem{gao15} L. Gao, F. Hou, and J. Huang. ``Providing Long-term Participation Incentive in Participatory Sensing''. \emph{IEEE International Conference on Computer Communications (INFOCOM)}, pp. 2803-2811, 2015.
\bibitem{restuccia16} F. Restuccia, S. K. Das, and J. Payton. ``Incentive Mechanisms for Participatory Sensing: Survey and Research Challenges''. \emph{ACM Transactions on Sensor Networks (TOSN)}, vol. 12, no. 2, 2016.
\bibitem{azadeh10} A. Azadeh, H. Mohamadlou, A. Pourahmad, S. Mohammadpour.``Modeling Road Traffic Accident Reporting System by Discrete Event Simulation: A Case Study of Iran's Road''. \emph{8th International Conference of Modeling and Simulation (MOSIM’10),  Hammamet - Tunisia}, pp. 1-10, 2010.

\bibitem{cardone14} G. Cardone, A. Cirri, A. Corradi, and L. Foschini. ``The Participact Mobile Crowd Sensing Living Lab: The Testbed for Smart Cities''. \emph{IEEE Communications Magazine}, vol. 52, no. 10, pp. 78-85, 2014.


\bibitem{gedeon15} J. Gedeon and I. Schweizer. 
``Understanding Spatial and Temporal Coverage in Participatory Sensor Networks'', \emph{IEEE International Conference on Local Computer Networks (LCN) Workshops}, pp. 699-707, 2015
\bibitem{zhang12} W. Zhang, B. Zhu, L. Zhang, J. Yuan, and I. You. ``Exploring Urban Dynamics based on Pervasive Sensing: Correlation Analysis of Traffic Density and Air Quality''. \emph{IEEE International Conference on Innovative Mobile and Internet Services in Ubiquitous Computing (IMIS)}, pp. 9-16, 2012.
\bibitem{foell13} S. Foell, G. Kortuem, R. Rawassizadeh, S. Phithakkitnukoon, M. Veloso, and C. Bento. ``Mining Temporal Patterns of Transport Behavior for Predicting Future Transport Usage''. \emph{ACM Conference on Ubiquitous Computing (UbiComp)}, pp. 1239-1248, 2013.
\bibitem{reddy08} S. Reddy, K. Shilton, J. Burke, D. Estrin, M. Hansen, and M. Srivastava. ``Evaluating Participation and Performance in Participatory Sensing''. \emph{International Workshop on Urban, Community, and Social Applications of Networked Sensing Systems (UrbanSense08)}, vol. 4, 2008.
\bibitem{tomasini17} M. Tomasini, B. Mahmood, F. Zambonelli, A. Brayner, and R. Menezes. ``On the Effect of Human Mobility to the Design of Metropolitan Mobile Opportunistic Networks of Sensors''. \emph{Elsevier Pervasive and Mobile Computing}, 2017.
\bibitem{zhao14} K. Zhao, M. Musolesi, P. Hui, W. Rao, and S. Tarkoma. ``Explaining the Power-Law Distribution of Human Mobility through Transportation Modality Decomposition'' \emph{Nature Scientific Reports}, 2015.
\end{thebibliography}
\vspace{-0.1in}
 
\end{document}